\documentclass[]{pasj02} 
\usepackage[switch,mathlines]{lineno} 
\usepackage{natbib} 
\usepackage[T1]{fontenc}
\usepackage{amsmath}
\usepackage{multirow}

\jyear{2024}
\Received{}
\Accepted{}

\newcommand{\sigmag}{\Sigma_\mathrm{g}}
\newcommand{\sigmad}{\Sigma_\mathrm{d}}
\newcommand{\del}{\partial}

\begin{document}

\title{Thermally driven spontaneous dust accumulation in the inner regions of protoplanetary disks}

\author{
Ryo \textsc{Kato}\altaffilmark{1}\altemailmark\orcid{0009-0007-6867-1459}\email{kato.r.ag@m.titech.ac.jp},
Takahiro \textsc{Ueda}\altaffilmark{2}\orcid{0000-0003-4902-222X},
Satoshi \textsc{Okuzumi}\altaffilmark{1}\orcid{0000-0002-1886-0880}
}

\altaffiltext{1}{Department of Earth and Planetary Sciences, Institute of Science Tokyo, Meguro, Tokyo 152-8551, Japan}
\altaffiltext{2}{Center for Astrophysics | Harvard \& Smithsonian, 60 Garden Street, Cambridge, MA 02138, USA}


\KeyWords{
protoplanetary disks ---
planets and satellites: formation --- planets and satellites: terrestrial planets} 

\maketitle

\begin{abstract}
In protoplanetary disks, the formation of planetesimals via streaming and/or gravitational instabilities requires regions with a locally enhanced dust-to-gas mass ratio. 
Conventionally, gas pressure maxima sustained by gas surface density maxima have been considered as the primary cause of such dust accumulation. 
However, the disk’s pressure structure depends not only on gas density but also on the temperature structure, which itself is influenced by the distribution of dust. 
In this study, we propose a novel mechanism for dust accumulation, which is driven by the coevolution of dust and disk temperature. 
In the inner disk region where the midplane temperature is primarily determined by the balance between viscous heating and radiative cooling, a perturbation in dust surface density distribution may affect radiative cooling efficiency, potentially producing a local maximum in the temperature and pressure profiles. 
To test this hypothesis, we perform coupled calculations of dust and disk temperature evolution, incorporating the advection, diffusion, coagulation, and fragmentation of dust particles along with viscous heating, radiative cooling, and radial thermal diffusion. 
Our results demonstrate that a pressure maximum formed by a perturbation in the dust surface density can spontaneously induce dust accumulation, even in the absence of a gas surface density maximum, under conditions where dust drift is significantly faster than diffusion and the thermal evolution occurs faster than the inward migration of dust.
This mechanism requires viscous heating to dominate disk heating, and typically occurs interior to the snow line. 
In this spontaneous dust trap, the dust-to-gas density ratio at the midplane can exceed unity, suggesting the potential for rocky planetesimal formation via streaming and gravitational instabilities.
\end{abstract}


\section{Introduction}\label{sec:intro}
Inner regions of protoplanetary disks are sites of rocky planetesimal formation.
The sticking growth of dust is inhibited by rapid radial drift caused by the negative pressure gradient along the radial direction
\citep[][]{Whipple1972,Adachi+1976,Weidenschilling1977}.
If there is a local maximum in the radial profile of disk gas pressure, the direction of dust radial drift reverses, leading to dust accumulation at the local pressure maximum
\citep[][]{Whipple1972,Brauer+2008,Kretke+2009,Pinilla+2012,Drazkowska+2013,Pinilla+2016,Ueda+2019,Ueda+2021}.
When the dust-to-gas density ratio at the midplane exceeds unity due to local dust accumulation, the gas is no longer able to diffuse dust particles in the vertical direction, leading to the formation of planetesimals through gravitational instability
\citep[e.g.,][]{Sekiya1998,Youdin+2002}.
A linear analysis \citep{Auffinger&Laibe2018} suggests that the streaming instability, which arises from the aerodynamic feedback from dust to gas \citep{Youdin＆Goodman2005,Johansen&Youdin2007}, can also operate at pressure maxima, but its mechanism is different from that of the streaming instability powered by the background pressure gradient. It is yet to be understood whether the instability at pressure maxima leads to nonlinear, strong dust clumping that results in planetesimal formation.

The gas pressure distribution in the disk depends on both the density and temperature distributions.
Therefore, comprehensive understanding of the evolution of the disk temperature structure is essential for studying planetesimal and planet formation.
The disk temperature structure is determined by the balance between cooling and heating.
The disks' primary cooling mechanism is radiative cooling via dust
\citep[e.g.,][]{D'Alessio+1999,Okuzumi+2022}, whereas the primary heating mechanisms are stellar irradiation and viscous heating \citep{Lynden-Bell+1974}.
Viscous heating refers to the process in which the gravitational potential energy released by viscous accretion of the gas heats the disk
\citep[e.g.,][]{Pringle1981,Nakamoto&Nakagawa1994,D'Alessio+1998}.
In the vicinity of the central star, viscous heating could be the dominant source of disk heating.

The midplane temperature determined by viscous heating depends on the vertical optical depth for dust thermal emission; a higher optical depth retains the released heat  for a longer time, leading to a higher midplane temperature \citep[e.g.,][]{Bell+1997,D'Alessio+1999}.
In this paper, we refer to this effect as the blanketing effect.
The optical depth is determined by the dust opacity and surface density. Dust growth leads to the depletion of small particles responsible for opacity \citep[e.g.,][]{Pollack+1994,Dullemond&Dominik2005}. Conversely, local dust accumulation increases the optical depth. These dust evolution processes alter the degree of the blanketing effect at different locations, thereby influencing the radial structure of the disk temperature and pressure \citep[][]{D'Alessio+2001,Oka+2011,Bitsch+2013,Bitsch+2014a}. 

Importantly, the change in pressure distribution caused by temperature variations due to dust evolution in turn alters the direction and speed of dust radial drift. This indicates that the evolution of dust and disk thermal structure is mutually coupled. The interplay between dust and temperature evolution is potentially crucial to planetesimal formation but has not been extensively investigated so far.

\begin{figure}
    \begin{center}
        \includegraphics[width = 80mm]{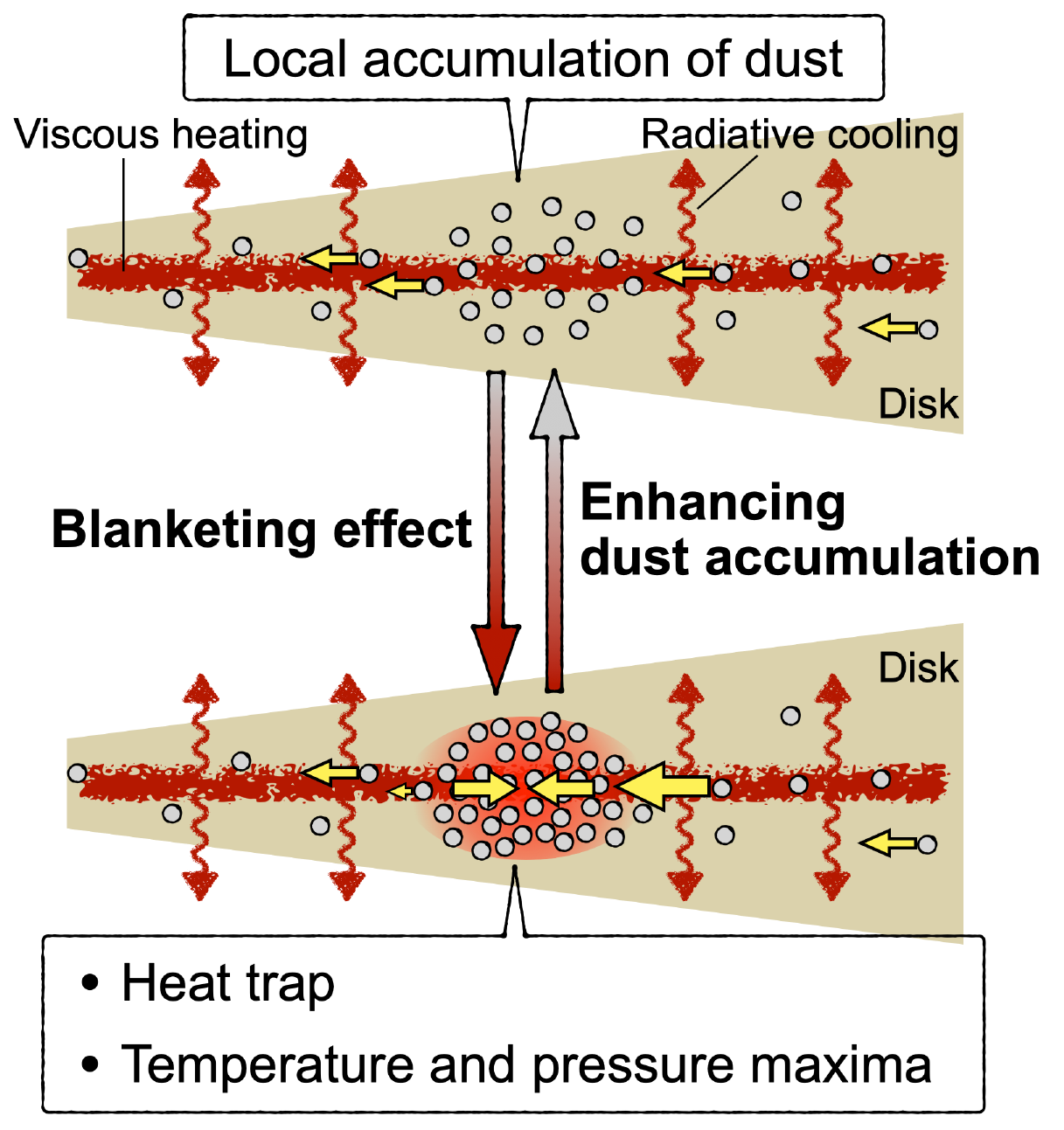}
    \end{center}
        \caption{
        Schematic of thermally driven spontaneous dust accumulation proposed in this study.
        
        {Alt text: A pair of cartoons depicting cross-sectional views of a disk, arranged vertically. The upper cartoon illustrates a local increase in dust density, while the lower cartoon depicts the corresponding rise in temperature and pressure. This feedback loop demonstrates how the increase in temperature and pressure further enhances the dust accumulation. }
        }
        \label{fig:Schematic}
\end{figure}

In this study, we propose a novel dust accumulation mechanism driven by pressure maxima that emerge from the coevolution of dust and disk temperature.
Figure~\ref{fig:Schematic} presents a schematic of the proposed mechanism.
Let us consider a situation where a small bump in the dust surface density distribution is formed by some mechanism. This bump then enhances the blanketing effect, increasing the temperature at this location. This temperature bump, in turn, produces a pressure bump, which slows down the drifting dust, resulting in a larger dust bump. 
We refer to this spontaneous dust accumulation as {\it thermally driven dust accumulation}. The aim of this study is to explore the conditions under which this mechanism operates.

The structure of this paper is as follows. 
Section \ref{sec:method} describes our numerical model for dust and gas disk evolution.
Section \ref{sec:results} presents the results of our numerical simulations and derives the conditions under which the dust accumulation mechanism proposed in this study occurs.
In section \ref{sec:discussion}, we discuss our proposed dust accumulation mechanism in the context of planetesimal formation, and assess the validity of our model.
A summary is presented in section \ref{sec:summary}.

\section{Method}~\label{sec:method}
In this section, we describe the model used to investigate the coevolution of dust and disk temperature structure. 
The calculations are conducted in one dimension along the radial direction. 
We simultaneously calculate the evolution of the surface densities of dust and gas, the size of dust particles, and the disk temperature.


\subsection{Dust and gas density evolution}
\label{sec:density_evo}
The evolution of the gas surface density $\sigmag$ is calculated by solving the equation of continuity, \begin{equation}
    \frac{\del \sigmag}{\del t}
    +
    \frac{1}{r} \frac{\del}{\del r}\left(r v_{{\rm g},r} \sigmag \right)
    = 0, 
    \label{eq:g_evo}
\end{equation}
where $r$ is the midplane distance from the central star, and $v_{{\rm g},r}$ is the radial velocity of gas, given by 
\begin{equation}
    v_{{\rm g},r} = 
    -\frac{3 \nu_{\rm t}}{r} \frac{\del \ln{r^{1/2} \nu_{\rm t} \sigmag}}{\del \ln{r}} ,  \label{eq:vg}
\end{equation}
with $\nu_{\rm t}$ being the gas turbulent viscosity coefficient.

We follow the evolution of dust surface density $\sigmad$ due to radial advection and diffusion by solving 
\begin{equation}
    \frac{\partial \sigmad}{\partial t}
    +
    \frac{1}{r}\frac{\del}{\del r}
    \left[
        r v_{{\rm d},r} \sigmad - 
        D_{\rm d} r\sigmag \frac{\del}{\del r}\left(
        \frac{\sigmad}{\sigmag} 
        \right)
    \right]
    = 0,    
    \label{eq:d_evo}
\end{equation}
where 
$D_{\rm d}$ is the dust diffusion coefficient.
We use the relation  \citep{Youdin&Lithwick2007} 
\begin{equation}
    D_{\rm d} = \frac{D_{\rm g}}{1+\rm St^2},
\end{equation}
where $D_{\rm g}$ is the gas diffusion coefficient and ${\rm St}$ is the Stokes number of the dust particles, defined as the product of the particles' stopping time and the disk's Keplerian frequency. We relate $D_{\rm g}$ to the viscosity as $D_{\rm g}=0.3 \nu_{\rm t}$ \citep{Okuzumi&Hirose2011}.

Neglecting the aerodynamic feedback from dust to gas, the dust radial velocity $v_{{\rm d},r}$ is given by 
\citep{Takeuchi&Lin2002}
\begin{equation}
    v_{{\rm d},r}
    = \frac{1}{1+{\rm St}^2} v_{{\rm g},r} + 
    \frac{2\rm St}{1 + {\rm St}^2}\Delta v_{{\rm g},\phi},
    \label{eq:vd}
\end{equation}
where 
\begin{equation}
    \Delta v_{{\rm g},\phi} = \frac{1}{2}\frac{c_{\rm s}^2}{v_{\rm K}}\frac{\del \ln{p_{\rm mid}}}{\del \ln{r}}
    \label{eq:vgphi}
\end{equation}
is the deviation of the gas azimuthal velocity from the Keplerian velocity $v_{\rm K}$ and $p_{\rm mid}$ is the midplane gas pressure. The gas pressure is expressed as $p=c_{\rm s}^2\rho_{\rm g}$, where $c_{\rm s}$ and $\rho_{\rm g}$ are the sound speed and gas density, respectively.
Dust backreaction on the gas is no longer negligible when the dust-to-gas density ratio at the midplane exceeds unity
\citep[e.g.,][]{Nakagawa+1986,Kretke+2009,Kanagawa+2017}.
We chose to neglect this effect as we are primarily interested in whether spontaneous dust accumulation operates and brings the dust-to-gas ratio above unity.
Subsequent dust evolution and planetesimal formation after the aerodynamic feedback becomes significant will be investigated in future work.

We approximate the particle Stokes number with its midplane value. Accounting for both the Epstein and Stokes drag regimes, we use (e.g., \citealt{Birnstiel+2010})
\begin{equation}
    {\rm St} = \frac{\pi}{2}\frac{\rho_{\rm {int}} a}{\sigmag}\max\left(1, \frac{4a}{9\lambda_{\rm mfp}}\right),
    \label{eq:St}
\end{equation}
where 
$\rho_{\rm int}$ and 
$a$ are the internal density and radius of the dust particle, respectively, and 
$\lambda_{\rm mfp}$ is the mean free path of the gas molecules written as 
$\lambda_{\rm mfp} = 
m_{\rm g}/
(\sigma_{\rm mol} \rho_{\rm g,mid})$,
with 
$m_{\rm g} = 3.82 \times 10^{-24}~\rm g$, 
$\sigma_{\rm mol} = 2.0 \times 10^{-15}~\rm cm^2$,
and 
$\rho_{\rm g,mid}$ 
being 
the mean molecular mass, 
the molecular collision cross section, and 
the midplane gas density, respectively.

Our model is essentially radially one-dimensional and does not require a detailed model for the vertical disk structure. For this reason, we approximate the  temperatures at all heights below the disk surface with the disk's midplane temperature $T_{\rm mid}$. 
The vertical gas density can then be analytically given by
\begin{equation}
    \rho_{\rm g}(r,z) = 
    \frac{\Sigma_{\rm g}(r)}{\sqrt{2 \pi}h_{\rm g}(r)}
    \exp{\left[-\frac{z^2}{2 h_{\rm g}(r)^2}\right]},   \label{eq:h_g}
\end{equation}
where $z$ is the height above the midplane and $h_{\rm g}$ is the gas scale height for a disk with the vertically uniform temperature of $T_{\rm mid}$. 
The gas scale height is related to the isothermal sound speed $c_{\rm s}$ at the midplane and Keplerian frequency $\Omega_{\rm K}$ as $h_{\rm g}=c_{\rm s}/\Omega_{\rm K}$.
The isothermal sound speed can be written as 
$c_{\rm s}=\sqrt{k_{\rm B}T_{\rm mid}/m_{\rm g}}$ with 
$k_{\rm B}$
being the Boltzmann constant.
In reality, an internally heated disk has a vertical temperature gradient. However a consistent calculation of the vertical temperature and gas density distributions requires numerical integration of the vertical hydrostatic equilibrium equation.
We avoid this complexity by using equation~\eqref{eq:h_g}.

For the gas turbulent viscosity, we use the $\alpha$-prescription \citep{Shakura+1973} 
\begin{equation}
    \nu_{\rm t} = \alpha c_{\rm s} h_{\rm g},
    \label{eq:nu}
\end{equation}
where $\alpha$ is the dimensionless viscosity.
{ In this study, we assume that disk turbulence is the source of the viscosity, neglecting radial angular momentum transport by laminar magnetic fields within the disk \citep[e.g.,][]{Bai&Stone2013, Lesur+2014,Gressel+2015}.
Disk accretion due to the vertical angular momentum removal by magnetic disk winds is also neglected. We note that magnetically driven disk accretion can cause the heating of disk interior, but its efficiency  depends on the height where Joule heat is released \citep{Hirose+Turner2011,Mori+2019,Mori+2021,Bethune+Latter2020,Kondo+2023}, which requires careful modeling.  }
To focus on the coevolution of disk temperature and dust, 
we do not consider the evolution of $\nu_{\rm t}$ in response to the temperature evolution.

\subsection{Dust size evolution} \label{sizeevo}
We consider dust size evolution due to coagulation and fragmentation.
To calculate the size evolution at low computational costs, we employ the single size approach of \citet{Sato+2016}. This approach assumes that the dust mass budget is dominated by particles of a representative mass $m_{\rm p}$, which depends on $t$ and $r$. In practice, $m_{\rm p}$ represents the mass of the largest particles at each radial position. We then calculate the evolution of $m_{\rm p}$ by solving
\begin{equation}
    \frac{\del m_{\rm p}}{\del t}
    +
    v_{\rm d,r}\frac{\del m_{\rm p}}{\del r}
    =
    \epsilon_{\rm grow} \frac{2 \sqrt{\pi} a^2 \Delta v_{\rm pp}}{h_{\rm d}} \Sigma_{\rm d},
\end{equation}
where $\epsilon_{\rm grow}$, $\Delta v_{\rm pp}$, and $h_{\rm d}$ are the sticking efficiency, collision velocity, and scale height of the mass-dominating particles, respectively.

The collision velocity $\Delta v_{\rm pp}$, accounts for the relative velocities induced by Brownian motion, radial and azimuthal drift, settling, and turbulence.
Following \cite{Sato+2016}, we take the Stokes numbers of two colliding particles to be ${\rm St_1} = {\rm St}$ and ${\rm St_2} = 0.5 {\rm St}$, where $\rm St$ is the Stokes number of the mass-dominating dust particles. 
The turbulence-induced velocity $\Delta v_{\rm t}$, which dominates $\Delta v_{\rm pp}$ in our simulations, is calculated from equations (26) and (28) of \cite{Ormel+2007}. It has simple asymptotic expressions
\begin{equation}
    \Delta v_{\rm t} \approx
    \begin{cases}
    \sqrt{\alpha}c_{\rm s}{\rm Re_t^{1/4}}|{\rm St_1 - St_2|}, & {\rm St_1} \ll 
    {\rm Re_t^{-1/2}}, \\
    \sqrt{2.3\alpha}c_{\rm s}{\rm St_1}^{1/2}, & 
    {\rm Re_t^{-1/2}}
    < {\rm St_1} \ll 1, 
    \end{cases}
\end{equation}
{ where ${\rm Re_t} = 2\nu_{\rm t}/(v_{\rm th}\lambda_{\rm mfp})$ is the turbulence Reynolds number, with $v_{\rm th} = \sqrt{8/\pi}c_{\rm s}$ being the thermal velocity of the gas.
The factor $\sqrt{2.3}$ \citep{Okuzumi&Tazaki2019} originates from equation (28) of \cite{Ormel+2007} with the assumption that ${\rm St}_1=0.5{\rm St}_2$.}

We model the sticking efficiency as \citep{Okuzumi+Hirose2012,Okuzumi+2016}
\begin{equation}
    \epsilon_{\rm grow} = \min \left[ 1, -\frac{\ln{(\Delta v_{\rm pp}/v_{\rm frag}})}{\ln{5}} \right],
\end{equation}
where $v_{\rm frag}$ is the threshold collision velocity above which the colliding particles fragment, i.e.,  $\epsilon_{\rm grow} < 0$. We assume that $v_{\rm frag}$ is constant throughout the 
{ computational domain}
and take it as a free parameter.
{ In real disks, $v_{\rm frag}$ may depend on the grains' chemical composition (see section~\ref{sec:param} for a discussion). However, we assume  $v_{\rm frag}$ to be radially constant, as we focus on the region inside the snow line, where the change of grain stickiness due to ice sublimation is negligible.}

Assuming the balance between the vertical settling and diffusion of dust particles, we express the dust scale height as \citep{1995Icar..114..237D,Youdin&Lithwick2007}
\begin{equation}
    h_{\rm d} = h_{\rm g} \left(1+\frac{\rm St}{\alpha_{\rm diff}}\frac{1+2\rm St}{1+\rm St}\right)^{-1/2},
\end{equation}
where $\alpha_{\rm diff} \equiv D_{\rm g}/(c_{\rm s}h_{\rm g}) = 0.3\alpha$ is the dimensionless turbulent diffusion coefficient.

\subsection{Temperature evolution} \label{tempevo}
We focus on the inner disk region where viscous heating is the dominant heat source.
To calculate the evolution of the midplane temperature, 
we use a vertically integrated energy equation that accounts for vertical
radiative cooling
and radial heat diffusion in addition to viscous heating
\citep{Watanabe+1990,Watanabe+2008,Okuzumi+2022},
\begin{equation}
    \begin{split}
    \frac{\gamma+1}{2(\gamma-1)}\frac{k_{\rm B}\sigmag}{m_{\rm g}}\frac{\del T_{\rm mid}}{\del t} 
    =
    &-2C\sigma_{\rm SB} T_{\rm mid}^4
    -
    \frac{1}{r}\frac{\del }{\del r}
    \left(r F_{\rm diff}\right) \\
    &+
    \frac{9}{4}\sigmag \nu_{\rm t} \Omega_{\rm K}^2,
    \end{split}
\end{equation}
where $\gamma = 1.4$ is the adiabatic index, 
$C$ is the dimensionless factor that corrects for the optical thickness and albedo effects on infrared radiation, 
$\sigma_{\rm SB}$ is the Stefan-Boltzmann constant, 
and  $F_{\rm diff}$ is the vertically integrated radial heat diffusion flux. 
The correction factor $C$ is given by \citep{Okuzumi+2022}
\begin{equation}
    C = \frac{4\sqrt{\epsilon}\tanh{(\tau_{\rm eff,mid})}}{\sqrt{3}+2\sqrt{\epsilon}\tanh{(\tau_{\rm eff,mid})}}
    \frac{1}{1+3\tau_{\rm R,mid}/4},
\end{equation}
where $\epsilon = \kappa_{\rm P}/\chi_{\rm R}$ is the ratio of the Planck-mean absorption opacity
\begin{equation}
    \kappa_{\rm P} = \frac{\pi}{\sigma_{\rm SB}T^4}\int_{0}^{\infty}\kappa_\nu B_\nu(T)d\nu
\end{equation}
to the Rosseland-mean extinction opacity
\begin{equation}
    \chi_{\rm R} = \frac{4\sigma_{\rm SB} T^3}{\pi \int_{0}^{\infty}\chi_\nu ^{-1}[dB_\nu(T)/dT]d\nu},
\end{equation}
with $\kappa_{\nu}$, 
$\chi_{\nu}$, and 
$B_{\nu}$
being 
the monochromatic absorption and extinction opacities at frequency $\nu$, and 
the Planck function,
respectively.
We express $\tau_{\rm eff,mid}$, the effective absorption optical depth to the midplane accounting for multiple scattering \citep{1979rpa..book.....R}, and $\tau_{\rm R,mid}$, the Rosseland-mean vertical optical depth to the midplane, as
\begin{equation}
    \tau_{\rm eff,mid} = \int_0^\infty \sqrt{3\epsilon}\chi_{\rm R}\rho_{\rm g}dz 
    = \int_0^\infty \sqrt{3\chi_{\rm R}\kappa_{\rm P}}\rho_{\rm g}dz
\end{equation}
and
\begin{equation}
    \tau_{\rm R,mid} = \int_0^\infty \chi_{\rm R}\rho_{\rm g} dz,
\end{equation}
respectively. 

The vertically integrated radial heat diffusion flux can be formally written as 
\begin{equation}
    F_{\rm diff} = \int_{-L_z/2}^{L_z/2} 
    \left[
    -\frac{4\sigma_{\rm SB}}{3\rho_{\rm g}\chi_{\rm R}}\frac{\del T(r,z)^4}{\del r}
    \right]
    dz.
    \label{eq:Fdiff}
\end{equation}
Here, we have applied the radiative diffusion approximation \citep[e.g.,][]{1979rpa..book.....R} to the region with a thickness of $L_z$, where the disk is radially optically thick to its own radiation.
Caution should be exercised when performing the integration in equation~\eqref{eq:Fdiff} because the factor  
$1/\rho_{\rm g}$ diverges toward larger $|z|$.
However, we expect that the radial temperature gradient 
should be small at large $|z|$ as a consequence of efficient radial heat diffusion. 
Since our model does not resolve the fully two-dimensional (radial and vertical) temperature distribution, we crudely approximate the integrand in equation~\eqref{eq:Fdiff} with its midplane value. This approximation yields
\begin{equation}
F_{\rm diff}  \approx
    - L_z
    \frac{4\sigma_{\rm SB}}{3\rho_{\rm g,mid}\chi_{\rm R}}
    \frac{\del T_{\rm mid}^4}{\del r}.
\end{equation}
In this study, we adopt $L_z = \sqrt{2\pi}h_{\rm g} \approx 2.5 h_{\rm g}$, which represents the effective thickness of the disk under the approximation that the disk has a vertically uniform density of $\rho_{\rm g,mid}= \Sigma_{\rm g}/L_z$ (see equation~\eqref{eq:h_g}).

The opacity per unit dust mass depends on the size of dust particles and the disk temperature.
We assume that, at each radial location, the dust size distribution ranges from 
$0.5~\mu\rm m$ to $a_{\rm max}$ with a power-law index of $-3.5$ 
\citep{Mathis+1977}, 
where $a_{\rm max}$ is computed from $m_{\rm p}$ as 
$a_{\rm max} = 
\left\{3 m_{\rm p}/(4 \pi \rho_{\rm int})\right\}^{1/3} $. 
The mean opacities are calculated as a function of $a_{\rm max}$ and $T_{\rm mid}$, assuming the optical constants for silicate dust \citep{Draine2003}, using the OpTool code \citep{Dominik+2021}. 

We neglect irradiation heating when determining the temperature. 
Neglecting stellar irradiation can be justified if $T_{\rm mid}$ from the simulation is higher than
\begin{equation}
    T_{\rm mid,irr} 
    \approx 
    150
    \left(
    \frac{r}{1~\rm au}
    \right)^{-{3}/{7}}
    ~\rm K,
    \label{eq:Ching_1997}
\end{equation}
which is the interior temperature of an optically thick, passively irradiated disk around a Sun-like T Tauri star \citep{Chiang_1997}.
In our simulations, this condition is met in the region of interest.

\subsection{Initial conditions}  \label{sec:ini}
We consider models with a bump structure in the initial gas or dust surface density profile. 
The background gas and dust surface densities are taken from on a steady, power-law accretion disk model. On top of the background disk, we introduce a bump centered at radius $r = r_{\rm cent}$.
In this paper, models with a bump in the initial gas surface density profile are referred to as gas bump models, whereas those with a bump in the initial dust surface density are referred to as dust bump models.
Below, we detail how the initial conditions are prepared.

First, the initial disk configuration is set so that the temperature at 1 au is the same in all models.
At the beginning of each simulation run, we set the initial midplane temperature $T_{{\rm mid},0}$ as 
\begin{equation}
    T_{{\rm mid},0} = T_{{\rm set}}
    \left(\frac{r}{1~\rm au}\right)^{-0.9}~{\rm K},
    \label{eq:ini_temp}
\end{equation}
where $T_{\rm set} \approx 250~\rm K$ is the initial midplane temperature at $1~\rm au$ of a disk in the equilibrium between viscous heating and radiative cooling, with a gas accretion rate of $\dot{M} = 10^{-8} M_{\odot}~\rm yr^{-1}$, $\alpha = 10^{-3}$, and $v_{\rm frag} = 10~\rm m~s^{-1}$.
The exponent of $r$ also comes from assumption of the balance between viscous heating and radiative cooling \citep[e.g.,][]{Oka+2011, Ida+2016}.
Then, assuming steady accretion \citep{Lynden-Bell+1974}, we determine the background gas surface density as
\begin{equation}
    \Sigma_{\rm g,bg} = \frac{\dot{M}}{3\pi\nu_{\rm t}}.
\end{equation}
{ We assign different values of $\alpha$ and $v_{\rm frag}$ to different simulation runs (see section~\ref{sec:param}).
Changing $\alpha$ or $v_{\rm frag}$ alters the initial surface density and dust particle size, which in turn affects the heating and cooling rates. In all models, we determine the value of $\dot{M}$ such that the temperature at $1~\rm au$ matches $T_{\rm set}$.}
We set the background dust surface density as 
\begin{equation}
    \Sigma_{\rm d,bg}=0.01\Sigma_{\rm g,bg}.
    \label{eq:d_bg}
\end{equation}
With these background surface density profiles, we let the midplane temperature and particle size evolve until equilibrium is reached, without radial transport of gas and dust and a gas or dust density bump. 

We then insert a bump in the gas or dust surface density profile. We define a function that characterizes the bump profile,
\begin{equation}
    f_{\rm bmp} = 1+A\exp\left[ - \frac{(r-r_{\rm cent})^2}{2\Delta r^2} \right],
\end{equation}
where $A$ and $\Delta r$ are the amplitude and width of the bump, respectively.
Gas bump models have the initial surface densities of $\sigmag = f_{\rm bmp}\Sigma_{\rm g,bg}$ and $\sigmad=\Sigma_{\rm d,bg}$, whereas dust bump models have $\sigmag=\Sigma_{\rm g,bg}$ and $\sigmad = f_{\rm bmp}\Sigma_{\rm d,bg}$.
All simulation models in this study adopt $A=1$,
$\Delta r = h_{\rm g}(r_{\rm cent})$, and 
$r_{\rm cent} = 1~\rm au$.
{ 
A smaller $A$ leads to slower dust accumulation, causing dust to drift toward the central star before significant accumulation occurs.
}

After the initial disk structure is determined, a full simulation is performed. 
{ Our computational domain ranges from $0.3~\rm au$ to $2~\rm au$ and is divided into 100 cells on a logarithmic scale.
We set the boundary conditions as follows. For the inner boundary, no inflow is assumed, while gas and dust are allowed to freely flow out of the computational domain. For the outer boundary, inflow is extrapolated based on the values of the cell adjacent to the boundary, and gas and dust are also allowed to freely flow out of the computational domain.}

\subsection{Parameter Choices} \label{sec:param}
{ We treat $\alpha$ and $v_{\rm frag}$ as free parameters since their values remain largely uncertain. Our choices of these values are shown in table~\ref{table:param}. 
}
{ 

We assume the viscosity parameter $\alpha$ to lie in the range  $10^{-4}$--$10^{-3}$. 
Observations suggest $\alpha \sim 10^{-4}$--$10^{-3}$ in the outer 10--100 au regions of protoplanetary disks \citep[e.g.,][]{Flaherty+2015, Dullemond+2018, Miotello+2023}. 
However, no strong observational constraints exist on $\alpha$ in the inner few au disk regions. Models predict that the inner disk regions may be susceptible to the convective overstability \citep{Klahr&Hubbard2014,Lyra2014} and zombie vortex instability \citep{Marcus+2013, Marcus+2015, Marcus+2016}, which could potentially drive turbulence with $\alpha \sim 10^{-5}$--$10^{-3}$ \citep[for a review, see][]{Lesur+2023}. 
Non-ideal MHD simulations by \cite{Bai2017} show that magnetic fields can radially transport angular momentum within a disk at an efficiency of $\alpha \sim 10^{-4}$--$10^{-3}$. However,  this effective viscosity includes contributions from laminar magnetic fields, which do not contribute to dust diffusion or collision velocities.

The fragmentation threshold $v_{\rm frag}$ is taken to be in the range 1--10 $\rm m~s^{-1}$. Previously, the fragmentation threshold for silicates inside the snow line was commonly assumed to be $\sim 1~\rm m~s^{-1}$ based on earlier laboratory experiments for aggregates composed of $1~\rm \mu m$-sized silica grains \citep[e.g.,][]{Blum&Wurm2000,Guttler+2010}.
However, more recent theoretical and experimental studies suggest that silicates may be stickier than previously expected \citep[e.g.,][]{Kimura+2015, Steinpilz+2019, Pillich+2021}.
Modeling of the hot FU Ori disk \citep{Liu+2021} and a massive protostellar disk \citep{Yamamuro+2023} provides further support for sticky silicates, with $v_{\rm frag}$ estimated to be $ \sim 10~\rm m~s^{-1}$.}



\begin{table}
\caption{Parameter sets adopted in our simulation models.}             
\label{table:param}      
\centering                          
\begin{tabular}{c c c c}        
\hline                 
\multicolumn{2}{c}{Model ID} & \multirow{2}{*}{$\alpha$} & \multirow{2}{*}{$v_{\rm frag}~[\rm m~s^{-1}]$} \\    
\cline{1-2}
Gas bump & Dust bump & \\
\hline                        
   GBa1e-3v10 & DBa1e-3v10 & $1\times10^{-3}$ & 10 \\      
   GBa1e-3v3 & DBa1e-3v3 & $1\times10^{-3}$ & 3  \\
   GBa1e-3v1 & DBa1e-3v1 & $1\times10^{-3}$ & 1  \\
   GBa3e-4v10 & DBa3e-4v10 & $3\times10^{-4}$ & 10 \\
   GBa3e-4v3 & DBa3e-4v3 & $3\times10^{-4}$ & 3  \\
   GBa3e-4v1 & DBa3e-4v1 & $3\times10^{-4}$ & 1  \\
   GBa1e-4v10 & DBa1e-4v10 & $1\times10^{-4}$ & 10 \\      
   GBa1e-4v3 & DBa1e-4v3 & $1\times10^{-4}$ & 3  \\
   GBa1e-3v1 & DBa1e-3v1 & $1\times10^{-4}$ & 1  \\
\hline                                   
\end{tabular}
\end{table}
\section{Results} \label{sec:results}
In this section, we present our simulation results to explore the condition under which thermally driven dust accumulation operates. 
The results from the gas and dust bump models are presented in sections~\ref{subsec:GB} and \ref{subsec:DB}, respectively.

\subsection{Gas bump models} \label{subsec:GB}
\begin{figure*}
\begin{center}
    \includegraphics[width = 160mm]{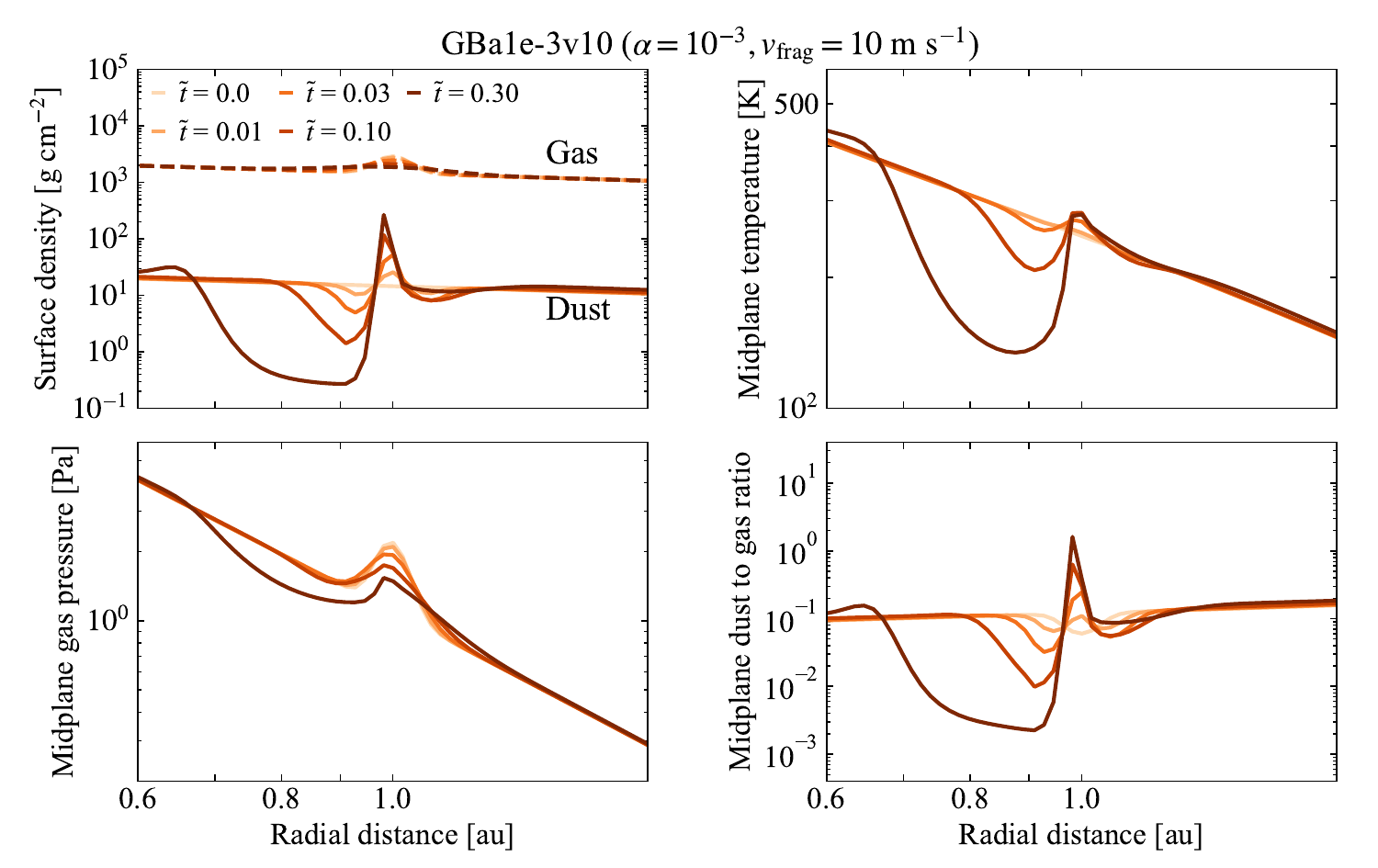}
    \end{center}
    \caption{
    Snapshots of the simulation result from gas bump model GBa1e-3v10.
    In the upper left panel, the solid and dashed lines represent the radial distributions of gas and dust surface densities $\Sigma_{\rm g}$ and $\Sigma_{\rm d}$, respectively.
    The upper right, lower left, and lower right panels show the temperature, gas pressure, and dust-to-gas density ratio, respectively, at the midplane. 
    The snapshots are taken at various normalized times
    $\tilde{t}\equiv t/t_{\rm drift}$, where $t_{\rm drift} \approx 1000~\rm yr$ for this run (see also equation~\eqref{eq:tdrift}).

    {Alt text: Graphs showing the radial profiles of various quantities at normalized times of 0, 0.01, 0.03, 0.1, and 0.3, obtained from the gas bump model with a turbulence strength of $10^{-3}$ and a fragmentation velocity of 10 meters per second.}
    }
    \label{fig:GB_alpha1e-3_vfrag10}
\end{figure*}

We select model GBa1e-3v10 ($\alpha = 10^{-3},~v_{\rm frag} = 10~\rm m~s^{-1}$) as a representative case in which thermally driven dust accumulation is observed.
Figure \ref{fig:GB_alpha1e-3_vfrag10} shows the evolution of the gas and dust surface densities, and the midplane values of the temperature, gas pressure, and dust-to-gas density ratio versus radial distance $r$ from this model.
Here, we label the snapshots by the normalized time $\tilde{t} \equiv t/t_{\rm drift}$, where  
\begin{equation}
    t_{\rm drift}
    =
    \left(\frac{r}{|2\Delta v_{{\rm g},\phi}{\rm St}|}\right)_{r = r_{\rm cent}}
\end{equation}
is the timescale for dust to drift from the position of the initial density bump, $r=r_{\rm cent}$, to the central star at the background drift velocity.
In this simulation, the maximum dust size is regulated by collisional fragmentation induced by turbulence, where the Stokes number satisfies $\Delta v_{\rm pp} \approx \sqrt{2.3\alpha \rm St }c_{\rm s} \approx v_{\rm frag}$ 
{ \citep{Birnstiel+2009, Okuzumi&Tazaki2019}.}
Therefore, $t_{\rm drift}$ can be estimated as
\begin{align}
   t_{\rm drift} &\approx
    \left(\frac{c_{\rm s}}{v_{\rm K}}\right)^{-2}
    \left|\frac{\del \ln{p_{\rm bg,mid}}}{\del \ln{r}}\right|^{-1}
    \frac{2.3\alpha c_{\rm s}^2}{v_{\rm frag}^2 \Omega_{\rm K}} \notag \\
    &\approx 1000\left(\frac{v_{\rm frag}}{10~\rm m~s^{-1}}\right)^{-2}
    \left( \frac{\alpha}{10^{-3}} \right)~\rm yr,
    \label{eq:tdrift}    
\end{align}
with $p_{\rm mid, bg}$ denoting the background pressure distribution at the midplane.
The final expression of equation \eqref{eq:tdrift} uses 
$r=r_{\rm cent}=1~\rm au$ and 
$\left|\del \ln{p_{\rm bg,mid}}/\del \ln{r}\right| \approx 2.55$.

\begin{figure*}
\begin{center}
    \includegraphics[width = 160mm]{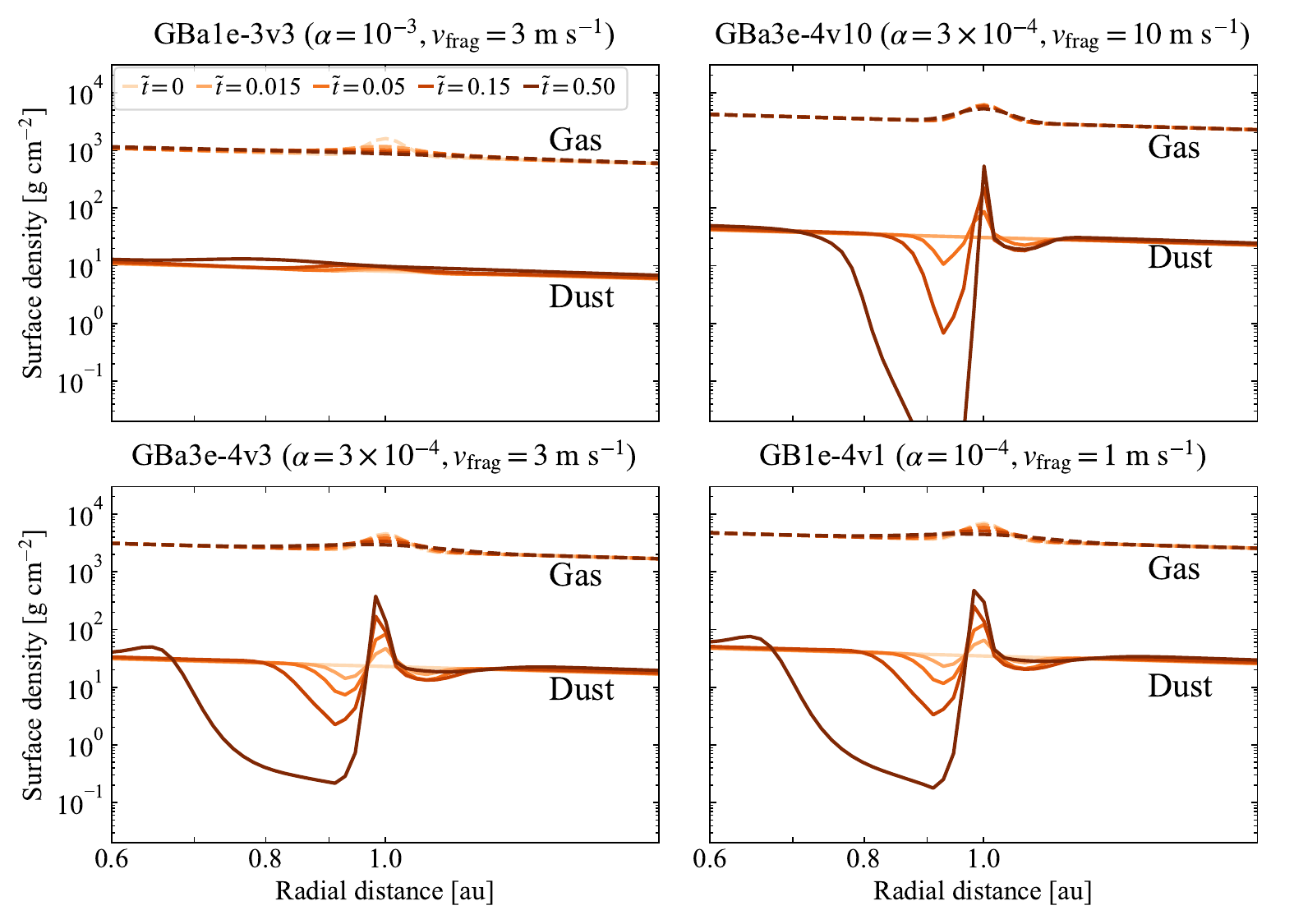}
    \end{center}
    \caption{
    Radial distributions of the gas and dust surface densities $\Sigma_{\rm g}$ and $\Sigma_{\rm d}$ (dashed and solid lines, respectively) at different normalized times $\tilde{t}\equiv t/t_{\rm drift}$ from gas bump models
    GBa1e-3v3 (upper left), 
    GBa3e-4v3 (upper right), 
    GBa3e-4v1 (lower left), and 
    GBa1e-4v1 (lower right).
    
    {Alt text: Graphs showing the radial distributions of gas and dust surface densities at normalized times of 0, 0.01, 0.03, 0.1, and 0.3. The combinations of turbulence strength and fragmentation velocity are as follows: $10^{-3}$ and 3 meters per second (upper left), $3\times 10^{-4}$ and 10 meters per second (upper right), $3\times 10^{-4}$ and 3 meters per second (lower left), and $10^{-4}$ and 1 meter per second (lower right).}
    }
    \label{fig:gusbump_4}
\end{figure*}
In figure~\ref{fig:GB_alpha1e-3_vfrag10}, one can see that dust initially accumulates toward the gas pressure maximum produced by the initial gas density bump.
Meanwhile, the gas density bump decays due to viscous spreading. 
However, we find that the dust accumulation is sustained even after the gas density bump has completely dissipated (see the snapshot at $\tilde{t} = 0.3$).
At this stage, the gas pressure distribution still has a local maximum at 1 au (see the lower left panel of figure~\ref{fig:GB_alpha1e-3_vfrag10}), but this pressure maximum is produced by the {\it temperature} maximum (see the upper right panel of figure~\ref{fig:GB_alpha1e-3_vfrag10}), not by the surface density maximum. This is why we refer to this mechanism as thermally driven dust accumulation.

The mechanism sustaining the dust and temperature maximum is as follows.
The dust surface density interior to the initial bump ($r < 1~\rm au$) decreases with time because the pressure bump traps inward drifting dust.
Accordingly, the temperature interior to the bump decreases because of the reduced blanketing effect.
In contrast, the temperature at the bump increases because of the enhanced blanketing effect.
These temperature variations create a local temperature maximum, which keeps trapping the dust even without the initial gas density bump. Since the dust accumulation is sustained, the resulting temperature bump is also sustained.

Whether the thermally driven dust accumulation occurs depends on the turbulence strength $\alpha$ and particle fragmentation velocity $v_{\rm frag}$. 
We demonstrate this in figure~\ref{fig:gusbump_4}, which present the results from four models with different values of $\alpha$ and $v_{\rm frag}$. 
Overall, we find that lower $\alpha$ and/or higher $v_{\rm frag}$ favors dust accumulation. 
This tendency can be readily understood as follows. Turbulence induces dust diffusion, which directly hinders local dust accumulation.
Turbulence also causes viscosity and acts to smear out the bump in the gas surface density.
Furthermore, collisional fragmentation by stronger turbulence results in smaller dust particles, making dust diffusion more efficient.
Consequently, strong turbulence generally suppresses the accumulation of dust at the gas bump.
On the other hand, a higher $v_{\rm frag}$ leads to larger particles with a higher drift velocity, facilitating dust accumulation toward the bump.
In all models resulting in thermally driven dust accumulation, the dust-to-gas density ratio at the midplane exceeds unity at $\tilde{t} \sim 0.3$, which is also confirmed in the lower right panel of figure~\ref{fig:GB_alpha1e-3_vfrag10}.

\begin{figure}
    \begin{center}
    \includegraphics[width = 65mm]{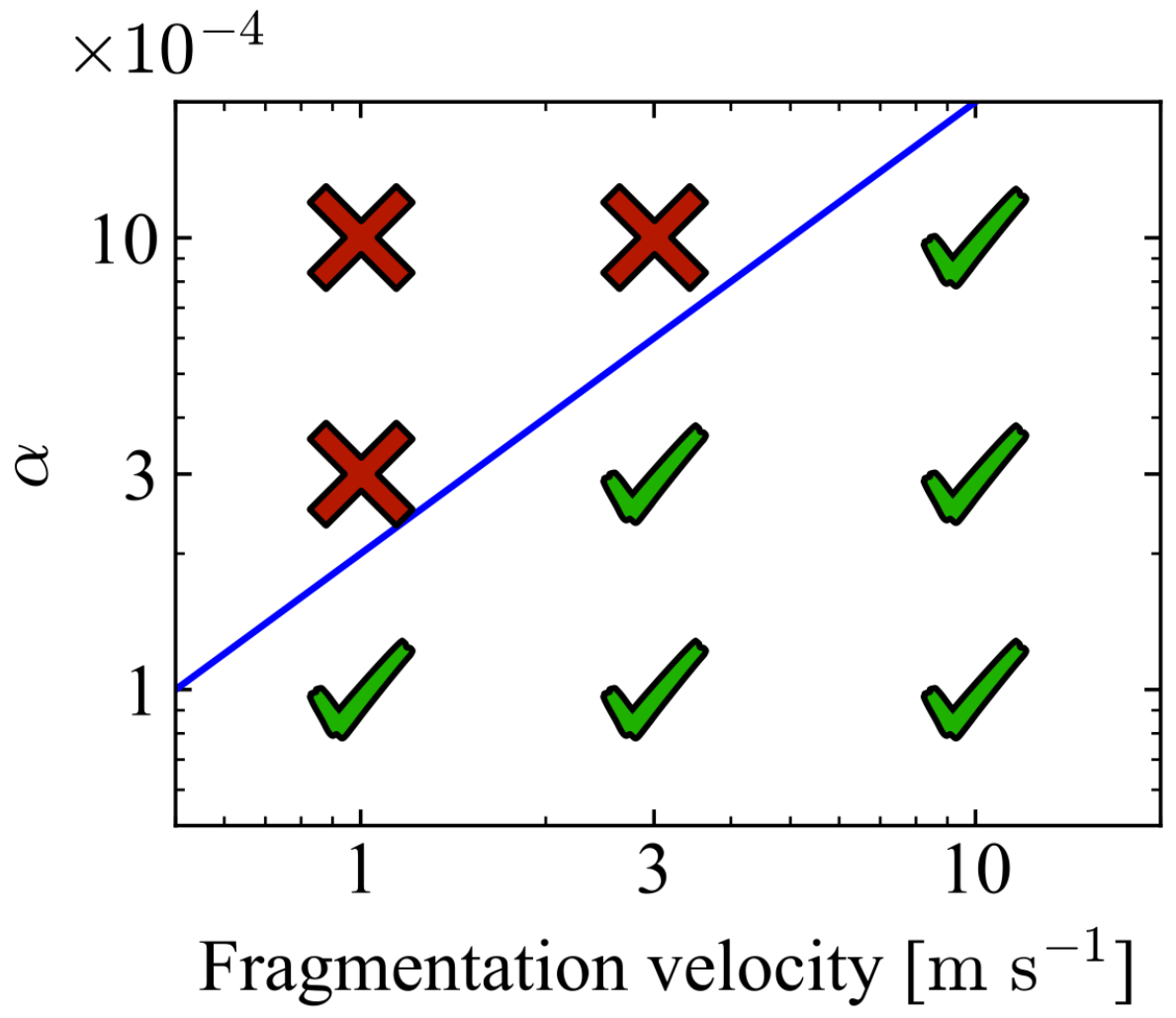}
    \end{center}
    \caption{
    Summary of gas bump models. 
    Check-marks indicate the parameter regions where the midplane dust-to-gas density ratio at the temperature bump exceeds unity, while cross-marks are used for regions where no local increase in dust surface density is observed.
    The solid line represents the threshold value of the turbulence intensity $\alpha$ for the thermally driven dust accumulation predicted by equation~\eqref{eq:threshold_alpha}.
    
    {Alt text: Graph showing the parameter space where thermally driven dust accumulation occurs in gas bump models, with fragmentation velocity on the horizontal axis and turbulence strength on the vertical axis.}
    }
    \label{fig:alpha_St}
\end{figure}
Figure~\ref{fig:alpha_St} indicates the parameter region where thermally driven dust accumulation is observed.
This figure suggests that there is an upper limit on $\alpha$ for dust accumulation, and that this upper limit increases with $v_{\rm frag}$.
Below, we analytically derive the condition for dust accumulation in terms of $\alpha$ and $v_{\rm frag}$.
We expect dust accumulation to occur when the timescale of gas diffusion over the bump width $ \approx h_{\rm g}$ is longer than that of dust drift over the same distance.
The diffusion and drift timescales for this lengthscale can be estimated as
\begin{equation}
    t_{{\rm g,diff},h_{\rm g}} = \frac{h_{\rm g}^2}{2D_{\rm g}}
    \approx \frac{1}{0.6 \alpha \Omega_{\rm K}}
    \label{eq:t_g_diff}
\end{equation}
and
\begin{equation}
    t_{{\rm drift},h_{\rm g}} = 
    \frac{h_{\rm g}}{2\Delta v_{{\rm g},\phi} \rm St} \approx
    \left(\frac{h_{\rm g}}{r}\right)^{-1} 
    \left| \frac{\del \ln{p_{\rm bg,mid}}}{\del \ln{r}} \right|^{-1}
    \frac{1}{\rm St \Omega_{\rm K}}
\end{equation}
Assuming $\rm{St}\ll1$, we can express the condition for the dust accumulation as
\begin{equation}
\begin{split}
    \frac{t_{{\rm g,diff},h_{\rm g}}}{t_{{\rm drift},h_{\rm g}}} 
    &\approx
    \frac{1}{0.6}
    \frac{h_{\rm g}}{r}
    \left|
        \frac{\del \ln{p_{\rm bg,mid}}}{\del \ln{r}}
    \right|
    \frac{\rm St}{\alpha} \\
    &\approx
    6
    \left(
        \frac{v_{\rm frag}}{10~\rm m~s^{-1}}
    \right)^2
    \left(
        \frac{\alpha}{10^{-3}}
    \right)^{-2}\gtrsim1,
    \label{eq:tdiff_tdrift}
\end{split}
\end{equation}
where we have used $r = r_{\rm cent} = 1~\rm au$, ${\rm St} \approx v_{\rm frag}^2/(2.3\alpha c_{\rm s}^2)$, and $\left|\del \ln{p_{\rm bg,mid}}/\del \ln{r}\right| \approx 2.55$.
Thus, the condition for dust accumulation
can be rewritten as
\begin{equation}
    \alpha
    \lesssim
    2\times 10^{-3}\left(\frac{v_{\rm frag}}{10~\rm m~s^{-1}}\right).
    \label{eq:threshold_alpha}
\end{equation}
The solid line in figure~\ref{fig:alpha_St} denotes the threshold value of $\alpha$ described by equation~\eqref{eq:threshold_alpha}, showing that equation~\eqref{eq:threshold_alpha} explains the simulation results well.

\subsection{Dust bump models} \label{subsec:DB}
In this subsection, we describe the simulation results of the models in which we assume a bump structure in the initial dust surface density profile.
Unlike gas bump models, dust bump models do not have an initial pressure maximum.
We investigate whether an initial dust bump can generate a pressure maximum and drive further dust accumulation.

\begin{figure}
    \begin{center}
        \includegraphics[width = 80mm]{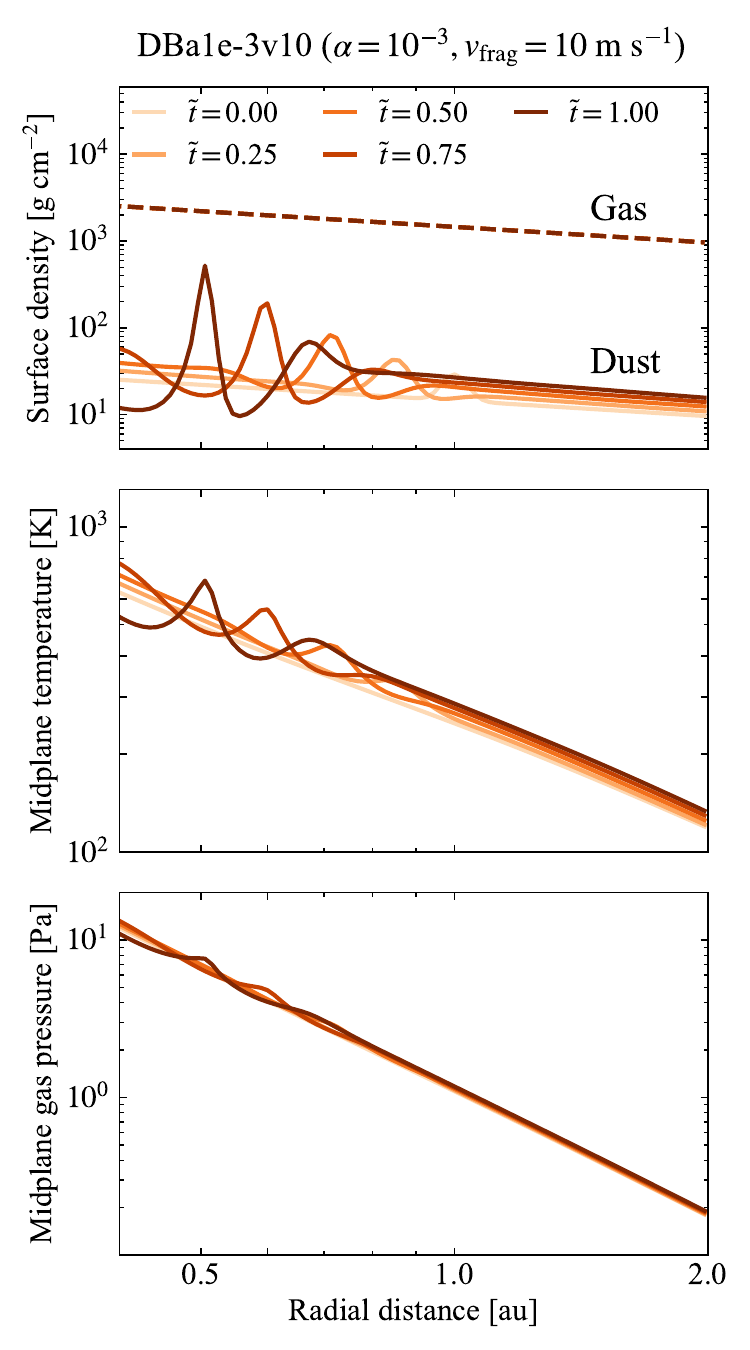}
        \end{center}
        \caption{
        Snapshots of the simulation result from dust bump model DBa1e-3v10 at different normalized times $\tilde{t} = t/t_{\rm drift}$.
        The panels show, from top to bottom, the gas and dust surface densities $\Sigma_{\rm g}$ and $\Sigma_{\rm d}$, midplane temperature $T_{\rm mid}$, and midplane gas pressure $p_{\rm mid}$. 

        {Alt text: Graphs showing the radial profiles of various quantities at normalized times of 0, 0.25, 0.5, 0.75, and 1, obtained from the dust bump model with a turbulence strength of $10^{-3}$ and a fragmentation velocity of 10 meters per second.}
        }
        \label{fig:DB_alpha3_vfrag10}
\end{figure}
\begin{figure}
    \begin{center}
    \includegraphics[width=80mm]{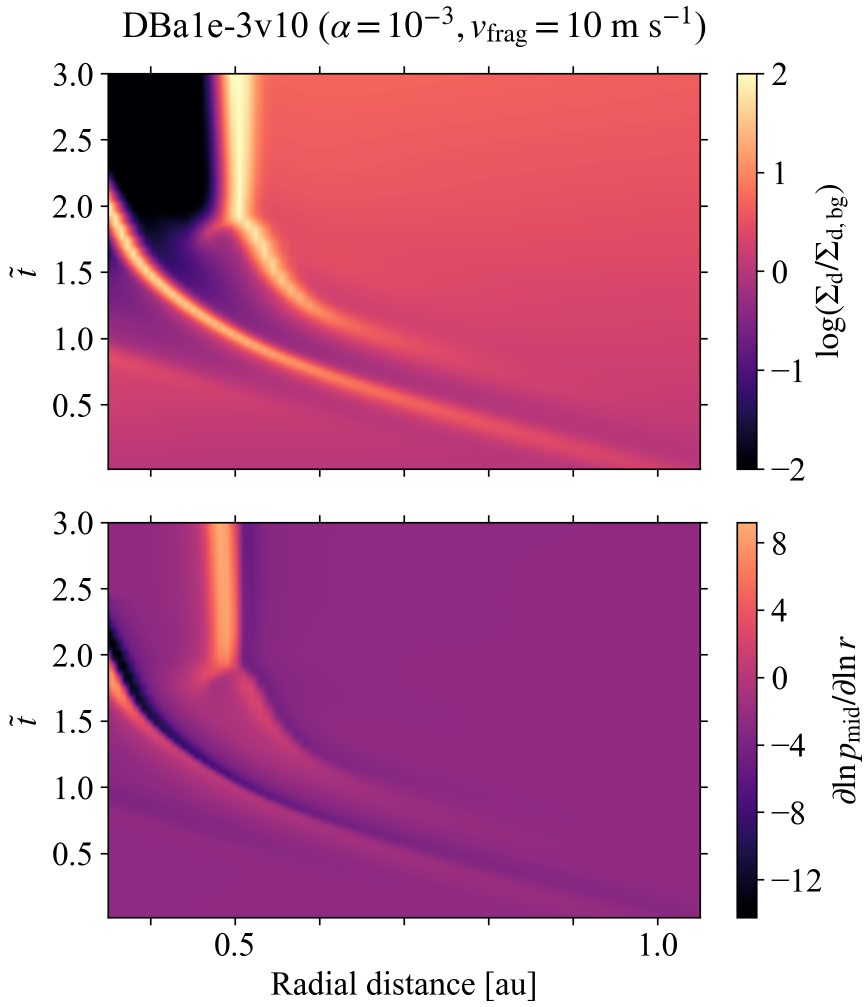}
    \end{center}
    \caption{Space time plots for the dust surface density $\sigmad$ normalized by background value $\Sigma_{\rm d,bg}$ (see equation~\eqref{eq:d_bg}), and the radial slope of the midplane gas pressure,  $\partial \ln{p_{\rm mid}}/\partial \ln{r}$, from dust bump model DBa1e-3v10.

    {Alt text: Two density plots showing the normalized dust surface density (top) and the radial slope of the midplane gas pressure (bottom) as function of the normalized time and radial distance. }
 }
    \label{fig:DBa1e-3v10_spacetime}
\end{figure}

We begin with model DBa1e-3v10 ($\alpha = 10^{-3}, v_{\rm frag} = 10~{\rm m~s^{-1}}$), which is a counterpart of the gas bump model GBa1e-3v10 presented at the beginning of subsection~\ref{subsec:GB}. This model serves as a representative case where an initial dust bump triggers thermally driven dust accumulation.
Figure~\ref{fig:DB_alpha3_vfrag10} shows the evolution of the gas and dust surface densities, the midplane temperature, and the midplane gas pressure from this model.
Due to the blanketing effect of the dust bump, the temperature increases, forming a temperature maximum near the bump.
In the bump's inner side, where the temperature gradient is positive, the negative gas pressure gradient is reduced, slowing down dust drift.
Conversely, in the outer side, the gas pressure gradient steepens, accelerating inward dust drift.
These lead to the growth of the dust bump.
The more pronounced dust bump further increases the local temperature and pressure. This positive feedback further promotes dust accumulation.
The growing dust bump migrates inward, but its inward migration is suppressed when the pressure gradient inside the bump inverts.
The spacetime plots in figure~\ref{fig:DBa1e-3v10_spacetime} show the evolution of the dust surface density $\Sigma_{\rm d}$ and midplane gas pressure gradient $\partial p_{\rm mid}/\partial \ln r$ as functions of $r$ and $\tilde{t}$ for model DBa1e-3v10.
In addition to the initial dust bump, the secondary dust bump emerges at $\tilde{t}\approx1$, which is caused by the density changes associated with the growth of the initial bump. 
Comparison between the plots for $\Sigma_{\rm d}$ and $\partial p_{\rm mid}/\partial \ln r$ shows that the inward migration of both the initial and secondary bumps is decelerated by the inversion of the pressure gradient. 
The secondary bump stops migrating completely as the positive pressure gradient on the bump's inner side becomes larger than the negative pressure gradient on the outer side.

\begin{figure*}
    \begin{center}
        \includegraphics[width = 160mm]{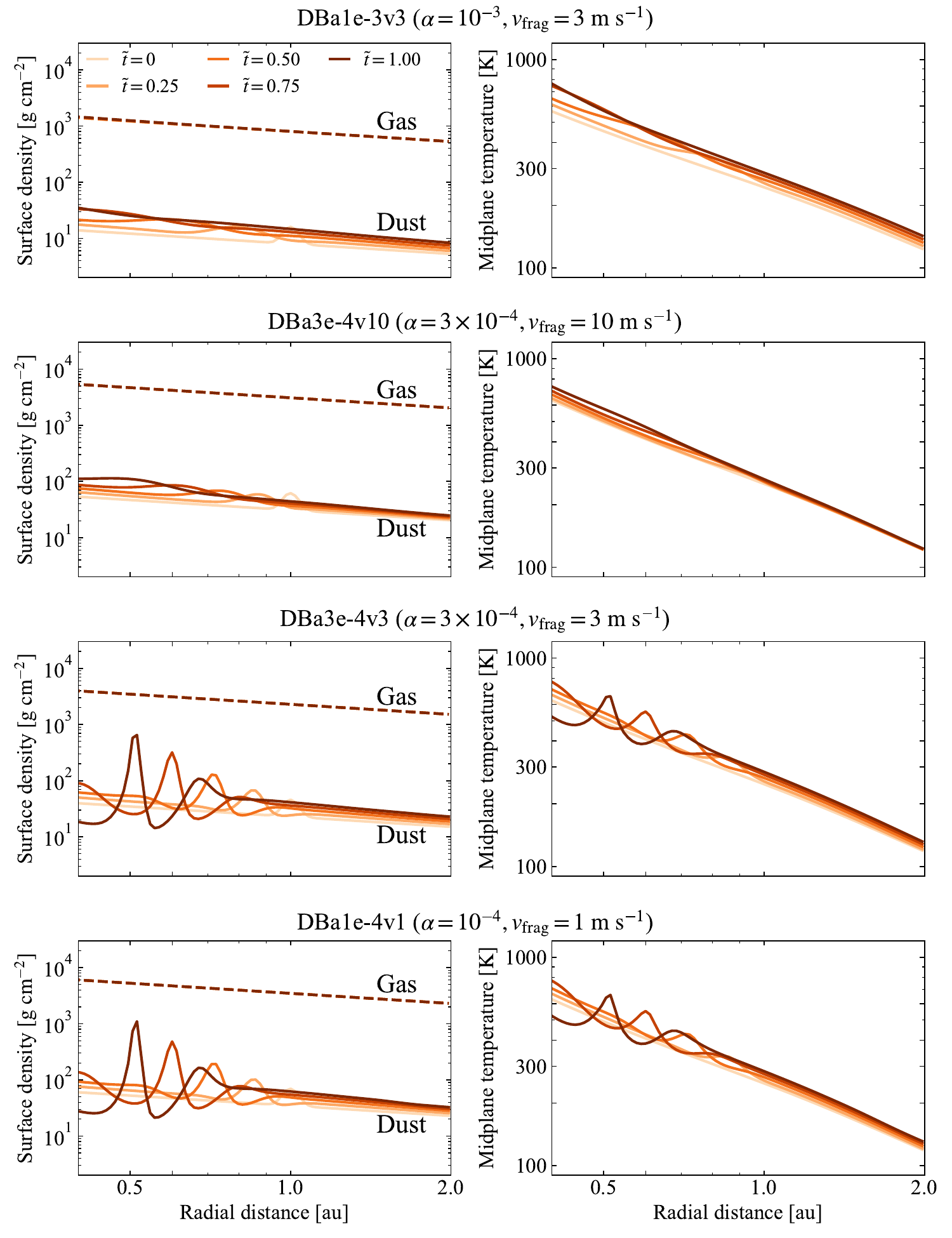}
        \end{center}
        \caption{Snapshots of the simulation results from dust bump models DBa1e-3v3, DBa3e-4v10, DBa3e-4v3, and 
        DBa1e-4v1 (from top to bottom).
        The left panels show the radial distributions of the gas and dust surface densities $\Sigma_{\rm g}$ and $\Sigma_{\rm d}$, whereas the right panels show those for the midplane temperature $T_{\rm mid}$.  
        
        {Alt text: Graphs showing the radial profiles of the gas and dust surface densities, and the midplane temperature at normalized times of 0, 0.25, 0.5, 0.75, and 1. The combinations of turbulence strength and fragmentation velocity are, from top to bottom, $10^{-3}$ and 3 meters per second, $3\times 10^{-4}$ and 10 meters per second, $3\times 10^{-4}$ and 3 meters per second, and $10^{-4}$ and 1 meter per second.}
        }
        \label{fig:4_2_DB}
\end{figure*}
In dust bump models, the occurrence of dust accumulation is governed by the timescales of thermal evolution, dust drift, and diffusion.
Figure~\ref{fig:4_2_DB} shows the simulation results of models 
DBa1e-3v3 ($\alpha = 10^{-3}, v_{\rm frag} = 3~{\rm m~s^{-1}}$),
DBa3e-4v10 ($\alpha = 3\times10^{-4}, v_{\rm frag} = 10~{\rm m~s^{-1}}$),
DBa3e-4v3 ($\alpha = 3\times10^{-4}, v_{\rm frag} = 3~{\rm m~s^{-1}}$), and 
DBa1e-4v1 ($\alpha = 10^{-4}, v_{\rm frag} = 1~{\rm m~s^{-1}}$).
In model { DBa1e-3v3}, 
the diffusion timescale is shorter than the timescale of dust drift across the bump, resulting in dust diffusion rather than accumulation.
In model DBa3e-4v10, no dust accumulation occurs despite that the drift timescale is shorter than the diffusion timescale.
This is because the drift timescale is significantly shorter than the thermal evolution timescale, causing the dust to drift inward before the temperature structure evolves.

\begin{figure}
    \begin{center}
    \includegraphics[width = 65mm]{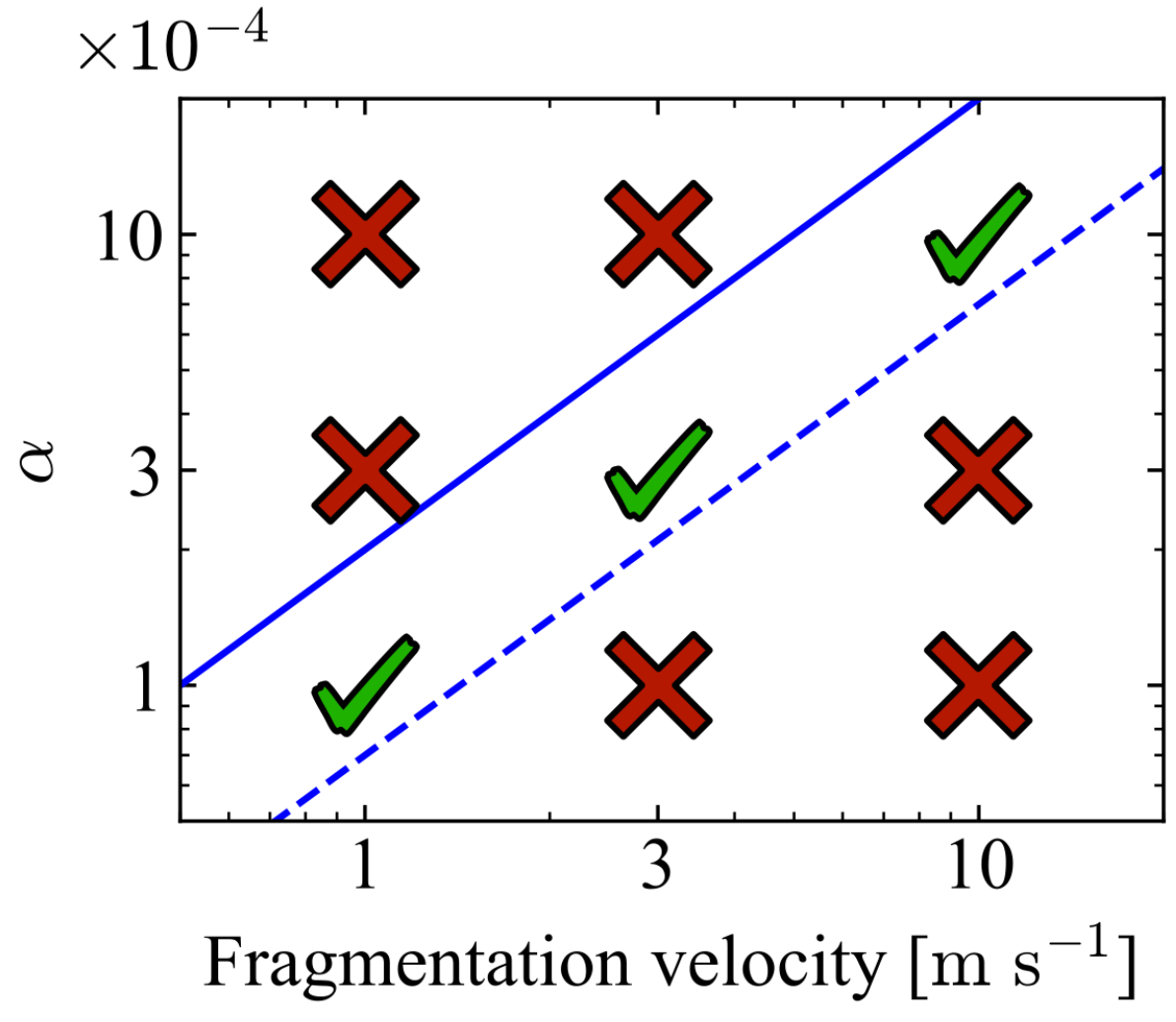}
    \end{center}
    \caption{
    Same as figure~\ref{fig:alpha_St}, but for dust bump models.
    The solid and dashed lines indicate the upper threshold value of $\alpha$ as described by equation~\eqref{eq:threshold_alpha}, and the lower threshold value as described by equation~\eqref{eq:threshold_alpha_2}, respectively.
    
    {Alt text: Graph showing the parameter space where thermally driven dust accumulation occurs in dust bump models, with fragmentation velocity on the horizontal axis and turbulence strength on the vertical axis.}
    }
    \label{fig:v_a_DB}
\end{figure}
In figure~\ref{fig:v_a_DB}, we show the parameter region where thermally driven dust accumulation occurs in the dust bump models.
Unlike gas bump models, dust accumulation in dust bump models has both an upper and a lower limit for its occurrence.
The upper limit is determined by the competition between the drift and diffusion timescales of dust, as in gas bump models (see equation \ref{eq:threshold_alpha}).
On the other hand, the lower limit is determined by the competition between the dust drift and viscous heating timescales, as shown below. For thermally driven dust accumulation to occur, 
 thermal evolution must occur faster than the dust drift. 
The timescale on which the midplane temperature varies from $T_{\rm mid}$ to  $T_{\rm mid} + \Delta T_{\rm mid}$  due to viscous heating can be estimated as
\begin{equation}
    t_{\rm heat} = 
    \frac{2(\gamma+1)}{9(\gamma-1)}
    \frac{k_{\rm B}\Delta T_{\rm mid}}{m_{\rm g}\nu_{\rm t}\Omega_{\rm K}^2}
    \approx
    \frac{4}{3\alpha \Omega_{\rm K}}
    \frac{\Delta T_{\rm mid}}{T_{\rm mid}}. 
\label{eq:t_heat}
\end{equation}
which increases as $\alpha$ decreases. 
Therefore, if $\alpha$ is too small, dust drifts toward the central star before the temperature bump required for dust accumulation develops (see the result of model DBa3e-4v10 shown in the second panel of figure~\ref{fig:4_2_DB}). 
Thus, in addition to equation \eqref{eq:threshold_alpha}, dust accumulation in dust bump models requires
\begin{equation}
\begin{split}
    \frac{t_{\rm heat}}{t_{{\rm drift},h_{\rm g}}} 
    &\approx
    \frac{4}{3}
    \frac{h_{\rm g}}{r}
    \left|
        \frac{\del \ln{p_{\rm bg,mid}}}{\del \ln{r}}
    \right|
    \frac{\rm St}{\alpha}
    \frac{\Delta T_{\rm mid}}{T_{{\rm mid}}} \\
    &\approx
    0.5
    \left(
    \frac{v_{\rm frag}}{10~\rm m~s^{-1}}
    \right)^2
    \left(
    \frac{\alpha}{10^{-3}}
    \right)^{-2}
    \lesssim 1, 
    \label{eq:theat_tdrift}
\end{split}
\end{equation}
where we have used $r = r_{\rm cent} = 1~\rm au$, ${\rm St} \approx v_{\rm frag}^2/(2.3\alpha c_{\rm s}^2)$, and $\left|\del \ln{p_{\rm bg,mid}}/\del \ln{r}\right| \approx 2.55$
, and 
$\Delta T_{\rm mid}/T_{\rm mid}\approx 0.1$.
From this, we obtain the lower limit of $\alpha$ for dust accumulation to occur in dust bump models as
\begin{equation}
    \alpha
    \gtrsim
    0.7\times 10^{-3}\left(\frac{v_{\rm frag}}{10~\rm m~s^{-1}}\right).
    \label{eq:threshold_alpha_2}
\end{equation}
In figure~\ref{fig:v_a_DB}, the upper solid line and the lower dashed line indicate boundaries of the parameter region expressed by equation \eqref{eq:threshold_alpha} and equation \eqref{eq:threshold_alpha_2}, respectively, showing that these boundaries well characterize the conditions for dust accumulation.

\section{Discussion} \label{sec:discussion}
\subsection{Implications for planetesimal formation}
The results of this study suggest that once a local maximum in the dust surface density is formed for some reason, the coevolution of dust and disk temperature can promote and sustain dust accumulation.
In previous studies, potential sites for dust accumulation due to gas pressure maxima have been attributed to gas surface density maxima located at sites such as the inner edge of the dead zone \citep[e.g.,][]{Dzyurkevich+2010, Flock+2017, Ueda+2019, Ueda+2021, Cecil&Flock2024} or the outer edges of gap structures \citep[e.g.,][]{Rice+2006, Zhu+2012}.
In our scenario, neither the sharp change in gas turbulence intensity required by the dead zone inner edge scenario nor the presence of a first-generation planet required by the gap outer edge scenario is necessary. 
Therefore, it is crucial to consider the coevolution of dust and disk temperature, incorporating the blanketing effect in planetesimal formation theories in regions where viscous heating is dominant source of heating.

Thermally driven dust accumulation requires viscous heating to be the dominant heat source within the disk, making it particularly crucial for rocky planetesimal formation in  inner disk regions.
In this study, we have assumed viscous accretion, in which the gravitational energy is primarily released at the disk midplane.
However, the accretion mechanism in real disks remains under debate.
The midplane heating may be suppressed if accretion is primarily driven by the magnetorotational instability in the upper layer of the disk \citep{Hirose+Turner2011} or by magneto-hydrodynamical disk winds \citep[e.g.,][]{Mori+2019, Mori+2021}.
However, accretion heating in magneto-hydrodynamically accreting disks can still influence the disk midplane temperature, depending on its ionization state and opacity \citep[e.g.,][]{Bethune+Latter2020,Kondo+2023}.
Future simulations incorporating these effects are crucial to understanding the formation of the building blocks of rocky planets.

\subsection{Impact of viscosity evolution on dust accumulation} \label{subsec:nuT}
In this study, we have neglected the response of gas viscosity to temperature evolution, by assuming that the viscosity is constant in time. 
In reality, the turbulent viscosity depends linearly with temperature as 
\begin{equation}
    \nu_{\rm t} = \alpha \frac{k_{\rm B} T_{\rm mid}}{m_{\rm g}}
    \frac{1}{\Omega_{\rm K}},
    \label{eq:nu_Tmid}
\end{equation}
where we have used equation \eqref{eq:nu}, $h_{\rm g} = c_{\rm s}/\Omega_{\rm K}$, and  
$c_{\rm s} = \sqrt{k_{\rm B}T_{\rm mid}/m_{\rm g}}$.
In the simulations presented in this study, we have fixed $T_{\rm mid}$ in equation~\eqref{eq:nu_Tmid} to be the initial midplane temperature profile.
In this subsection, we briefly examine how temperature-dependent viscosity evolution impacts dust accumulation.

\begin{figure*}
    \begin{center}
        \includegraphics[width = 150mm]{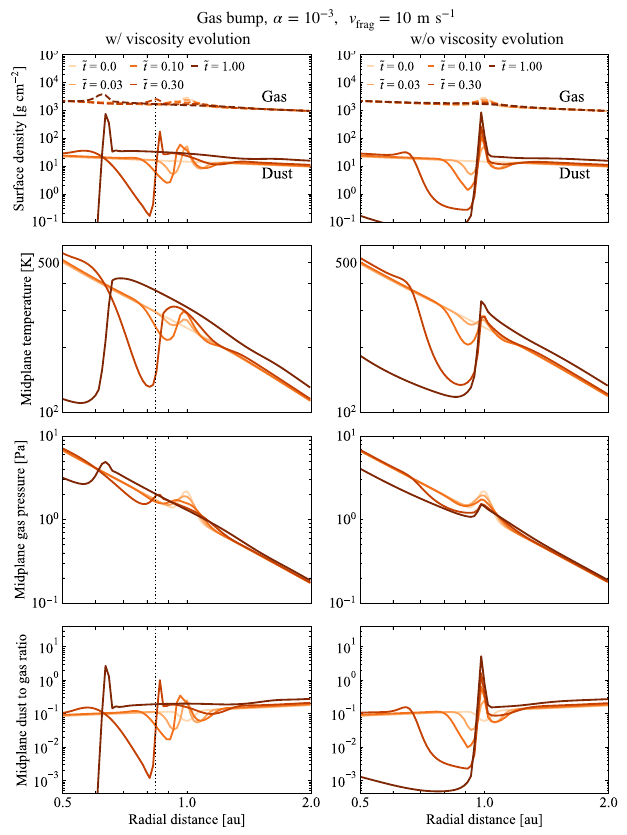}
        \end{center}
        \caption{
        { Snapshots at different normalized times $\tilde{t}=t/t_{\rm drift}$ comparing the results of a model incorporating the evolution of gas viscosity $\nu_{\rm t}$ in the gas bump model GBa1e-3v10 (left column) with the original GBa1e-3v10 results (right column). The panels show, from top to bottom, the gas and dust surface densities $\sigmag$ and $\sigmad$, midplane temperature $T_{\rm mid}$, midplane gas pressure $p_{\rm mid}$, and midplane dust-to-gas density ratio. In the left column, the dotted lines mark $r = 0.84~\rm au$, where a new gas surface density maximum emerges at $\tilde{t}\approx0.3$ (see the top panel).}
        
        {Alt text: Graphs showing the radial profiles of various quantities at normalized times of 0, 0.03, 0.1, 0.3, and 1. The combination of turbulence strength and fragmentation velocity is $10^{-3}$ and 10 meters per second.}
        }
        \label{fig:nuon_GB_alpha3_vfrag10}
\end{figure*}
Here, we rerun model GBa1e-3v10 ($\alpha = 10^{-3}, v_{\rm frag} = 10~{\rm m~s^{-1}}$), allowing $T_{\rm mid}$ in equation~\eqref{eq:nu_Tmid}
to evolve. Figure \ref{fig:nuon_GB_alpha3_vfrag10} shows 
{ 
the evolution of the gas and dust surface densities, and the midplane values of the temperature, gas pressure, and dust-to-gas density ratio from model GBa1e-3v10 with and without the evolution of $\nu_{\rm t}$ in response to $T_{\rm mid}$.}
{ From the left column of figure~\ref{fig:nuon_GB_alpha3_vfrag10}, it is observed that in the simulation with viscosity evolution, although the initial gas bump dissipates due to diffusion, }
a new gas bump emerges at $\tilde{t} = t/t_{\rm drift} = 0.3$ and $r \approx 0.84~ \rm au$, which is 
slightly inward of the temperature maximum formed by dust accumulation.
The emergence of the new gas bump is due to the dependence of the viscosity-driven gas velocity on the temperature gradient, $v_{{\rm g},r} \propto - \partial \nu_{\rm t}/\partial r \propto -\partial T_{\rm mid}/\partial r$ (see equation \ref{eq:vg}). The viscosity-driven inward velocity $-v_{{\rm g},r}$ is enhanced interior to the temperature maximum because of the positive temperature gradient, causing a pileup of gas just inside the temperature maximum.
Since gas pressure depends on both the temperature and gas density, the newly formed gas bump shifts the pressure bump inward.
Consequently, the dust bump, as well as the temperature bump, migrates inward, in contrast to the run without viscosity evolution {(see the right column of figure~\ref{fig:nuon_GB_alpha3_vfrag10}).} 
Unlike gas bump models with no viscosity evolution, the dust bump keeps migrating inward. However, the midplane dust-to-gas ratio exceeds unity well before the dust falls toward the central star (see the snapshot at $\tilde{t}=0.3$ in the fourth panel of { the left column of} figure~\ref{fig:nuon_GB_alpha3_vfrag10}), indicating that it could trigger planetesimal formation in the disk.
The gas bump induced by the viscosity evolution primarily contributes to the inward migration of the dust accumulation position, while the dust accumulation itself is directly driven by the contribution of the temperature maximum caused by the blanketing effect.


\begin{figure*}
    \begin{center}
        \includegraphics[width = 150mm]{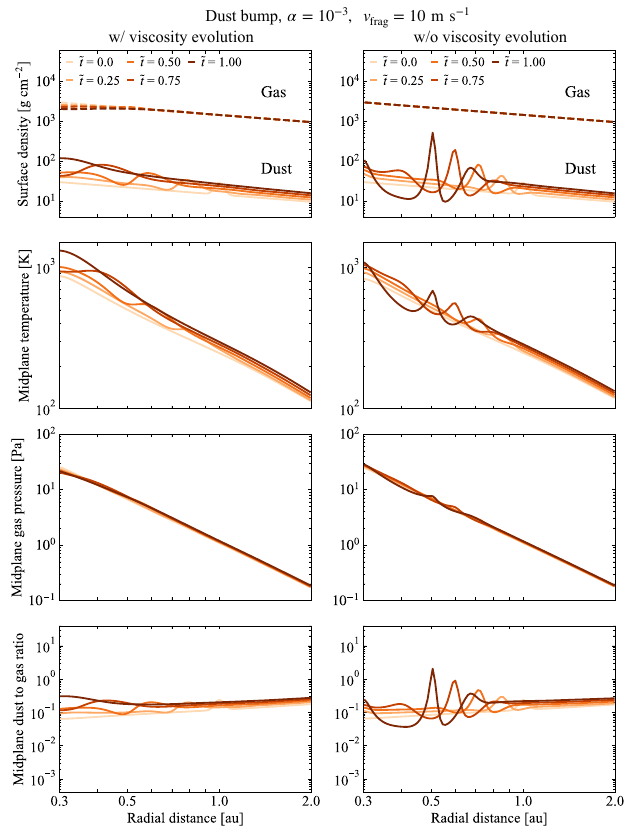}
        \end{center}
        \caption{
        { Same as figure~\ref{fig:nuon_GB_alpha3_vfrag10}, but for the dust bump model. The simulation parameters ($\alpha$, $v_{\rm frag}$) and initial conditions are the same as those for dust bump model DBa1e-3v10.
        } 
        
        {Alt text: Graphs showing the radial profiles of various quantities at normalized times of 0, 0.25, 0.5, 0.75, and 1. The combination of turbulence strength and fragmentation velocity is $10^{-3}$ and 10 meters per second.}
        }
        \label{fig:nuon_DB_alpha3_vfrag10}
\end{figure*}
\begin{figure}
    \begin{center}
        \includegraphics[width = 75mm]{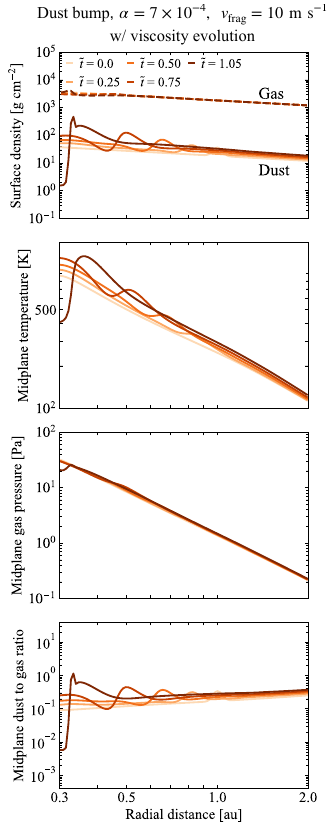}
        \end{center}
        \caption{
        { Same as the left column of figure~\ref{fig:nuon_DB_alpha3_vfrag10}, but with $\alpha=7\times10^{-4}$ and $v_{\rm frag}=10~\rm m~s^{-1}$.}
        
        {Alt text: Graphs showing the radial profiles of various quantities at normalized times of 0, 0.25, 0.5, 0.75, and 1. The combination of turbulence strength and fragmentation velocity is $7\times10^{-4}$ and 10 meters per second.}
        }
        \label{fig:nuon_DB_alpha7_vfrag10}
\end{figure}

{ Figure~\ref{fig:nuon_DB_alpha3_vfrag10} shows the results of dust bump model DBa1e-3v10 with and without the evolution of $\nu_{\rm t}$ with $T_{\rm mid}$. In the case with viscosity evolution (left column), unlike the case without it (right column), the initial dust bump does not grow sufficiently to form a pressure bump, and there is almost no increase in the local dust-to-gas density ratio at the midplane. The reason for this is as follows. The blanketing effect due to the dust bump induces a positive temperature gradient just interior to the bump (see the snapshots at $\tilde{t}=0.25$ in the second panels). In the case with viscosity evolution, this positive temperature gradient accelerates  gas accretion. As a result, the dust inside the bump is dragged inward, preventing dust accumulation.
However, thermally driven dust accumulation occurs 
if the dust drift relative to the gas (second term on the right-hand side of equation~\ref{eq:vg}) is faster than inward accretion of dust with gas (first term on the right-hand side of equation~\ref{eq:vg}). Such conditions can be achieved by either a smaller $\alpha$ or a larger $v_{\rm frag}$. 
Figure~\ref{fig:nuon_DB_alpha7_vfrag10} shows the results from a dust bump model with $\alpha = 7\times10^{-4}$, $v_{\rm frag}=10~\rm m~s^{-1}$, slightly smaller $\alpha$ than that of model GBa1e-3v10, with viscosity evolution. In this case, the smaller $\alpha$ reduces the co-accretion velocity of dust with gas, allowing the dust to accumulate and the dust-to-gas density ratio at the midplane to exceed unity before the dust falls toward the central star (see the snapshot at $\tilde{t}=1.05$ in the fourth panel of figure~\ref{fig:nuon_DB_alpha7_vfrag10}).} 

{ In summary, we have shown that temperature-dependent viscosity modifies thermally driven dust accumulation, yet the changes in the temperature structure induced by the blanketing effect can still drive dust accumulation. 
In gas bump models, the temperature dependence of viscosity yields a new gas surface density bump that migrates inward. This bump causes the accumulated dust to migrate inward as well. However, the midplane dust-to-gas density ratio exceeds unity well before the dust falls to the central star. 
In dust bump models, although the increased gas accretion velocity suppresses dust accumulation, thermally driven dust accumulation can still operate when 
dust drift is faster than inward co-accretion with gas.

In reality, whether the viscosity is temperature-dependent would depend on its underlying driving mechanism, which is poorly constrained. In turbulence driven by local hydrodynamical instabilities (e.g., convective overstability and zombie vortex instabilities discussed in section~\ref{sec:param}), the turbulent velocity should scale with the local sound speed, and therefore the resulting viscosity (proportional to the turbulence velocity squared) should scale with the local temperature. 
In contrast, when magnetohydrodynamics is the dominant driver of disk accretion and heating, the equivalent viscosity scales with the Maxwell stress, which depends on the magnetic field strength but not on temperature \citep[in other words, $\alpha$ scales with the plasma beta; see, e.g.,][]{Bai&Stone2011,Okuzumi&Hirose2011}. Furthermore, laminar magnetic fields, which are neglected in this study, can drive gas accretion but do not diffuse dust, potentially promoting dust accumulation. 
However, magnetically driven accretion does not always cause efficient heating of the disk interior, particularly when the associated Joule heating occurs predominantly near the relatively well-ionized disk surface \citep{Hirose+Turner2011,Mori+2019,Mori+2021,Bethune+Latter2020,Kondo+2023}. Self-consistent modeling of Joule heating with magnetically driven accretion is needed to elucidate if thermally driven dust accumulation operates in magnetically accreting disks.}

\section{Summary} \label{sec:summary}
We investigate how dust evolution and disk temperature evolution coevolve in a viscous accretion disk and how this coevolution affects dust accumulation.
To achieve this, we simultaneously calculate the evolution of dust and disk temperature, taking into account the blanketing effect.
As initial conditions, we consider gas bump models, which have a peak in the gas surface density distribution, and dust bump models, which have a peak in the dust surface density distribution.

In gas bump models { (section~\ref{subsec:GB})}, dust accumulates at the initial pressure maximum when the drift timescale of dust is significantly shorter than the diffusion timescale over the bump width { (equation~\ref{eq:tdiff_tdrift})}.
Furthermore, even after the gas surface density maximum diffuses, the blanketing effect helps maintain the pressure maximum, promoting further dust accumulation.
{ The conditions under which thermally driven dust accumulation operates in gas bump models are summarized in figure~\ref{fig:alpha_St}.}
In dust bump models { (section~\ref{subsec:DB})}, dust accumulation occurs when the drift timescale of dust is shorter than the diffusion timescale { (equation~\ref{eq:tdiff_tdrift})} and longer than the thermal evolution timescale { (equation~\ref{eq:theat_tdrift})}. 
Unlike in gas bump models, there is no initial pressure maximum in dust bump models.
Therefore, a local maximum in temperature and pressure profiles need to be formed due to the blanketing effect.
Even if the drift timescale of dust is shorter than the diffusion timescale, dust will drift inward before the temperature structure can change due to the blanketing effect if the thermal evolution timescale is significantly longer than the drift timescale.
In this case, thermally driven dust accumulation is inhibited.
{ The conditions under which thermally driven dust accumulation operates in dust bump models are summarized in figure~\ref{fig:v_a_DB}.}

This study indicates that if perturbations occur in the surface density of dust or gas due to certain factors, the coevolution of dust and disk temperature can lead to the self-formation and maintenance of gas pressure maxima, which may result in dust accumulation.
This suggests that planetesimal formation can occur even without a local maxima in the gas surface density profile, such as those caused by sharp changes in the strength of gas turbulence at inner edges of dead-zones or at outer edge of gaps created by first-generation planets.

{ Accounting for the evolution of viscosity as a function of disk temperature modifies the dust accumulation mechanisms, but the blanketing effect can still trigger thermally driven dust accumulation (section~\ref{subsec:nuT}). Temperature-dependent viscosity accelerates the inward gas velocity just interior to the temperature bump, hindering dust accumulation. However, at the same time, this difference in the inward gas velocity across the temperature bump forms a gas bump, which acts to facilitate dust accumulation. 
In reality, whether viscosity varies with disk temperature changes would depend on its underlying mechanism. When laminar magnetic fields are the dominant driver of disk accretion, the corresponding effective $\alpha$ does not contribute to dust diffusion, potentially facilitating dust accumulation. However, in magnetically accreting disks, midplane heating may be suppressed, which may hinder thermally driven dust accumulation. Simulations that self-consistently incorporate Joule heating and magnetically driven accretion are required to verify whether thermally driven dust accumulation can operate in magnetically accreting disks.}


\begin{ack}
The authors thank Mario Flock, Michael Cecil, and David Melon Fuksman for helpful discussions on thermal structure of disks, and Alexandros Ziampras for constructive comments. The authors also thank the referee, Daniel Carrera, for his constructive comments on the temperature dependence of viscosity.
\end{ack}

\section*{Funding}
This work was supported by JSPS KAKENHI Grant Numbers JP20H00205, JP23H00143, and JP23K25923.
T.U. acknowledges the support of the JSPS overseas research fellowship. 

\bibliographystyle{apj}
\bibliography{ThermallyDriven}

\end{document}